\documentclass[prl,twocolumn]{revtex4-1}

\usepackage{blindtext}

\usepackage{graphicx}  
\usepackage{dcolumn}   
\usepackage{bm}        
\usepackage{amssymb}   
\usepackage{epsfig}
\usepackage{amsmath} 
\usepackage{color}
\usepackage{natbib}

\begin{document}
\title{{Modelling the environmental dependence of the growth rate}}
\author{Ixandra Achitouv$^{1,2}$ \& Yan-Chuan Cai$^{3}$}
\email{ixandra.achitouv@obspm.fr}
\affiliation{$^1$Laboratoire Univers et Th\'eories (LUTh), UMR 8102 CNRS, Observatoire de Paris, Universit\'e Paris Diderot, 5 Place Jules Janssen, 92190 Meudon, France\\
$^2$APC, Univ Paris Diderot, CNRS/IN2P3, CEA/lrfu, Obs de Paris, Sorbonne Paris Cité, France\\
$^3$Institute for Astronomy, University of Edinburgh, Royal Observatory, Blackford Hill, Edinburgh, EH9 3HJ , UK}

\date{\today}

\begin{abstract} 
The growth rate of cosmic structure is a powerful cosmological probe for extracting information on the gravitational interactions and dark energy.
In the late time Universe, the growth rate becomes non-linear and is usually probed by measuring the two point statistics of galaxy clustering in redshift space up to a limited scale, retaining the constraint on the linear growth rate $f$. In this letter, we present an alternative method to analyse the growth of structure in terms of local densities, i.e. $f(\Delta)$. Using N-body simulations, we measure the function of $f(\Delta)$ and show that structure grows faster in high density regions and slower in low density regions. We demonstrate that $f(\Delta)$ can be modelled using a log-normal Monte Carlo Random Walk approach, which provides a means to extract cosmological information from $f(\Delta)$. We discuss prospects for applying this approach to galaxy surveys.

\end{abstract}

\maketitle
\noindent The growth rate of cosmic structure contains important information on the matter-energy content of the Universe and the gravitational interactions that shape the cosmic web. A powerful way to extract this information is to use redshift-space distortions (RSD) in galaxy clustering (e.g.~\citep{Peacock2001,Tegmark2006,Reid2012,delaTorre2013}), or in the cross-correlation between clusters/voids and galaxies ~\citep{Zu2013,HamausSDSS,AH_Viper2016,AB_voids2016}. However, when using RSD, among other cosmological probes, we are limited by the accuracy of our model to reproduce complex patterns in the galaxy clustering on small scales. Hence we are often forced to throw away data in the non-linear regime in order to extract unbiased cosmological information,  in this case, the linear growth rate (e.g.~\citep{Beutler6dF,AB_voids2016, Cai16,Nadathur17,AchitouvPRD17}). One way to overcome this issue is to use perturbative approaches to model the global clustering in the quasi-nonlinear regime down to a certain small scale where models break down. While non-linear modelling allows us to extract an unbiased value of the linear growth rate, in principle, two point statistics such as the correlation function is sensitive to the variance of the field. Applying them to a non-linear field will not be able to extract all the information. This is because a non-linear density field is usually non-Gaussian, and can not be fully characterised by its variance. One can use higher order statistics such as 3-point or 4-point correlation functions to regain the information beyond the variance, but this is currently computationally expensive. 

\medskip
\noindent In this study, we propose a different approach towards the same problem: instead of measuring the globally averaged linear growth rate $f$ at different scales by forward modelling the non-linear growth of the matter power spectrum/correlation function, we accept that the growth of structure depends on local densities and aim to model this dependency, i.e. $f(\Delta)$, where $\Delta=\frac{\rho}{\bar\rho}-1$ is the local density contrast. To do this, we analyse the growth rate using numerical simulations in and around overdense and underdense regions and show how it can be predicted as a function of local density and for a given cosmology. This prediction relies on log-normal Monte Carlo Random Walks, a method introduced in~\citep{IA2016}. We find that our model is successful in tracking the evolution of the growth rate at different local density environments. This, in principle, provides an independent method to extract cosmological information from the quasi-linear and non-linear regime. Our method of understanding the non-linear growth is in the same spirit of modelling the distribution of densities within spheres \cite{Bernardeau2015, Uhlemann2016, Codis2016}, density split statistics \cite{Friedrich2017, Gruen17} and the modelling of the non-linear aspect of the BAO \citep{AB_BAO,Neyrinck2018}. A more complete study will be presented in a companion paper. 

\begin{figure}
\begin{center}
\includegraphics[scale=0.45]{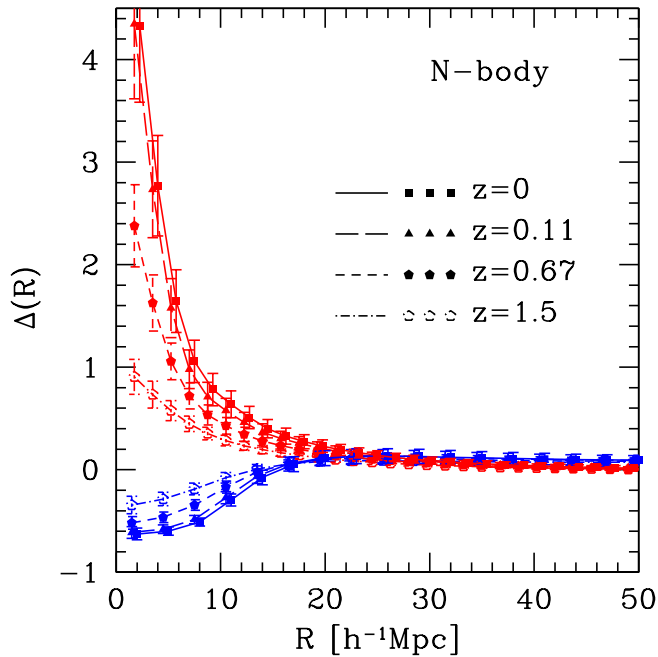}
\caption{Cummulative matter density profiles around overdense (red) and underdense (blue) regions, measured from $\Lambda$CDM N-body simulations at different redshifts indicated by the legend. }\label{Fig1}
\end{center}
\end{figure}

\medskip
\noindent We perform our analysis using N-body simulations from the DEUS consortium. These are described in~\citep{Alimi2010,Courtin2011,Rasera2010} and are publicly available. These simulations are run in a $\Lambda$CDM model with the WMAP-5yr cosmology \cite{wmap5} with ($w=-1;\Omega_m=0.26;\sigma_8=0.79)$. They have box-lengths of 648$h^{-1}$Mpc with $1024^3$ particles. They were generated using the RAMSES code~\citep{Teyssier2002}; halos were found using an FoF finder with the link-length $b=0.2$, \cite{Davisetal85} and cover a range of masses $M\sim [10^{12}-10^{15}] h^{-1}\rm M_{\odot}$.

\medskip
\noindent We first identify regions of different density contrast $\Delta(R)$ (i.e. environement) in the simulations, where $R$ is the radius of the region. We follow the method presented in~\citep{IA2016} to identify low density regions, i.e. voids. This algorithm imposes density thresholds at the radius of our choice, therefore allowing flexibility to represent a large variety of void profiles. Here we choose $R=20 h^{-1} \rm Mpc$ and the same criteria for the voids as the ones used in~\citep{IA2016}. This fiducial size is statistically motivated to obtain enough non-overlapping voids. We run this void finder on the halo catalog and we find $\sim 2300$ void centers. We measure the dark matter density profiles around our selected void centres to avoid complication due to the halo bias. 
{We select the overdense regions by randomly sampling positions of dark matter particles belonging to halos above the mass resolution at $z=0$, until we reach the same number of overdensities as the number of voids, to make sure that these two samples have similar noise properties. Keeping the same comoving coordinates for the under/overdense regions fixed, we measure the evolution of the density profiles at  redshifts: $z=\{0.00;0.05;0.11; 0.67,1.50\}$  (corresponding to scale factors $a=\{1.00,0.95,0.90,0.60,0.40\}$ respectively).

\medskip
\noindent In Fig.~\ref{Fig1} we show the mean matter density profiles of these over/underdense patches (dots) at different redshifts. 
From these profiles we measure numerically the growth rate within the radius $R$ at $a=0.95$ using {three consecutive snapshots at $a=\{1.00,0.95,0.90\}$} by computing 
\begin{equation}
f(R)\equiv \frac{d\ln \Delta(R)}{d\ln a},\label{fdef}
\end{equation}
where $\Delta(R)$ is the cumulative density contrast. The centers of our under/overdense regions are kept unchanged at different epochs, so we are actually tracking the evolution of the growth rate within each small  ``island Universe'', characterised by its density.

\begin{figure}
\begin{center}
\includegraphics[scale=0.45]{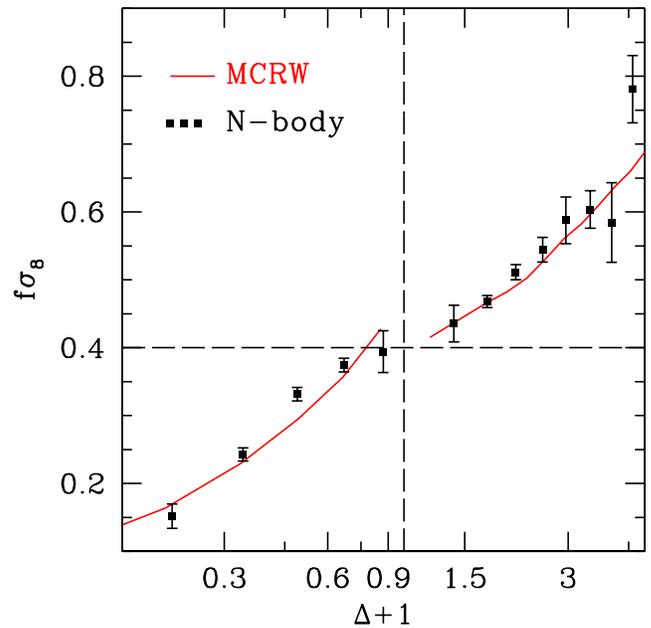}
\caption{{Growth rate parameter $f\sigma_8$ measured around a range of regions characterised by their density contrasts $\Delta$ from a $\Lambda$CDM N-body simulation (dots with errors) and our model prediction from the log-normal MCRW approach (red curve). The horizontal dashed line shows the linear expectation. The vertical dashed line indicates the means density of the universe. } }\label{Fig2}
\end{center}
\end{figure}

\medskip
\noindent The resulting growth rates are shown in Fig.~\ref{Fig2} by the black data points. We bin up the $f$ values according to their local density $\Delta$ to show the values of $f \sigma_8$ as a function of the local density, where $\sigma_8=0.79$ is a constant. {Note that the density contrasts of different scale $R$ may end up in the same bin of $\Delta$. In this sense, the behaviour of $f(\Delta)$ is no longer an explicit function of scale $R$, but depends solely on the local density $\Delta$, which could be contributed by perturbations of different scales.} The error bars correspond to the standard deviation computed from the mean measurements of 64 sub-cubes of length $162h^{-1}$Mpc. The horizontal dashed line corresponds to the linear growth rate. 
Although Eq.~\ref{fdef} has a logarithmic divergence for $\Delta \rightarrow 0$, we can see how the growth rate varies compared to the linear one, indicated by the horizontal dashed line in Fig.~\ref{Fig2}. The growth of structure slows down in low density regions and speed up in high density regions. While $f(R)$ is expected to reach the linear value on large scales, the growth rate in terms of $\Delta$ only crosses over the linear value when $|\Delta|$ is small. It is therefore important to go beyond a single value of the linear $f$ by modelling the whole spectrum of $f(\Delta)$ to extract cosmological information effectively. 

\medskip
\noindent In general, we expect the overall averaged growth rate on small scales to be higher than in linear theory. This is because the amplitudes of the late-time matter power spectrum/correlation function tend to be higher than the linear version on small scales. These higher amplitudes must arise from a higher growth rate. This suggests that the larger/smaller growth rate in over/underdense regions seen in Fig.~\ref{Fig2} does not exactly cancel out for the global average on these scales. In fact, the matter power spectrum/correlation function on small scales is dominated by high density regions. Therefore, the branch of the curve with $\Delta > 0$ shown in Fig.~\ref{Fig2} contributes more to the global averaged growth rate than the $\Delta < 0$ branch does. This is also consistent with the general trend of the inferred value of the growth rate from redshift space distortions (e.g.~\citep{Beutler6dF,AB_voids2016, Cai16}). The models used to analyse the redshift space distortions measurements are considered within a fitting range that excludes the small-scale clustering.
For instance in~\citep{Beutler6dF}, the authors infer the linear growth rate using galaxy-galaxy redshift space distortions with a cutting scale along the line-of-sight $> 10 h^{-1} \rm Mpc$. 

\medskip
\noindent While the small scale information with expected higher growth rate is usually disregarded due to the limitations of models, the main idea of our study is to provide a description for the growth rate on these non-linear scales. To develop this model, we could try to reproduce the density profiles we show in Fig.\ref{Fig1}. For instance using the well-known Zeldovitch approximation~\citep{Zeldovitch70}, which links the initial density profiles $\Delta(a_{\rm ini})$ to a later time $\Delta(a)$ assuming no shell-crossing and mass conservation (e.g.~\citep{deFromont2017}). These approximations, as well as the spherical evolution (e.g.~\citep{Gunn&Gott1973}~\citep{Bernardeauspherical}) have been investigated in the literature and recently the authors of~\citep{deFromont2017} have found that both Zeldovitch and spherical evolution lead to a similar evolution of an initially spherical density perturbation, which is in very good agreement with N-body simulations in some special cases (e.g. voids that are compensated, $\Delta(R=R_v)>0$, where $R_v$ is the radius of a void).
However, these two methods that describe the non-linear evolution have one main disadvantage: they require as an input the initial density perturbation $\Delta(R,a_{\rm ini})$. The evolution of this initial density profile becoming non-linear at the late time, a small modification in the initial input can lead to very different predictions of $\Delta(R,a=1)$. This makes it very difficult, from an observational point of view, to probe precisely the initial densities {and connect them to cosmologies}, although recent developments have been made through probing projected void density profiles (e.g.~\citep{Gruen17}). 

\medskip 
\noindent In this study we adopt an approach that has the advantage of not requiring {the initial condition of density profiles}. Instead of modelling the global non-linear evolution of densities in terms of scales, as done in perturbation theories, we generalise the non-linear evolution of the growth rate $f$ as function of the local density, which is equivalent to having a model for $f(\Delta)$, where $\Delta(R)$ is the value of the density contrast within the radius $R$. Our approach is referred to as \textit{log-normal Monte Carlo Random Walks} (MCRW) and has been developped in~\citep{IA2016}. It relies on the empirical observation that the late time probability density function (PDF) of the galaxies (hence the dark matter density fluctuations), is well-described by a log-normal  PDF (e.g.~\citep{Hamilton1985, Bouchetetal1993, Kofmanetal1994}). This has been confirmed by several studies using N-body simulations (e.g.~\citep{Coles&Jones1991,Kofmanetal1994,Taylor&Watts2000}) even in the highly non-linear regime (down to $R\sim 2h^{-1} \rm Mpc$ for $\Lambda$CDM ~\citep{KayoTaruyaSuto2001}). Using this log-normal (LN) assumption, the author~\citep{IA2016} has generated a set of log-normal \textit{Monte Carlo Random Walks}. These walks are ensembles of density contrast vectors $\Delta_{LN}(R)$, that are numerically generated from a log-normal distribution, and aim to describe the density contrasts around random positions in the late-time Universe. The starting point of this method uses the framework of the excursion set theory~\citep{Bondetal1991}: for Gaussian initial density perturbations, the evolution of the density contrast, smoothed on a scale $R$ and at a random position (e.g. $\textbf{x}=0$), is 

\begin{equation}
\frac{\partial \Delta (R,\textbf{x}=0)}{\partial R}=\int \frac{d^3 k}{2\pi^3}\;  \tilde{\delta}_k \; \frac{\partial \tilde{W}(k,R)}{\partial R} \label{EqLang}
\end{equation}

\noindent where $\tilde{\delta}_k$ and $\tilde{W}(k,R)$ are the Fourier transforms of the density fluctuation, and the filter function (top-hat in real space), respectively. For Gaussian initial conditions, $\tilde{\delta}_k$ satisfies  $\left<\tilde{\delta}_k \tilde{\delta}_k'\right> \equiv \delta_{\rm D}(k-k')P_{\rm lin}(k)$, where $P_{\rm lin}(k)$ is the linear matter power spectrum. For each initial realization of the density fluctuations $\tilde{\delta}_k$, the stochastic differential Eq.~\ref{EqLang} can be solved numerically assuming that $\Delta(R\rightarrow \infty)=0$ (e.g.~\citep{Bondetal1991}). Hence we have a discrete set of values $\{\Delta(R_1),\Delta(R_2),..\Delta(R_N)\}$ at each smoothing scale $\{R_1,R_2,...R_N\}$, that is by definition one random walk. Repeating this process for a large number of initial density fluctuations allows us to generate Gaussian random walks. In order to describe the later-time non-linear density fluctuation, we follow~\citep{IA2016}, and take the log-normal transformation of each Gaussian random walk using 

\begin{equation}
\begin{split}
&\Delta_{\rm LN}+1=\frac{1}{\sqrt{1+\sigma_{\rm NL}^{2}(R)}}\times \\
&\exp\left(\frac{\Delta}{\sigma_{\rm lin}(R)}\sqrt{\ln(1+\sigma_{\rm NL}^{2}(R))}\right),\label{mapping}
\end{split}
\end{equation}

\noindent with 

\begin{subequations}
\begin{align}
       \sigma^{2}_{\rm lin}(R)\equiv\frac{1}{2\pi^2}\int P_{\rm lin}(k) \tilde{W}^2(k,R) k^2 dk \label{sigmaLi} \\  
       \sigma^{2}_{\rm NL}(R)\equiv\frac{1}{2\pi^2}\int P_{\rm NL}(k) \tilde{W}^2(k,R) k^2 dk  \label{sigmaNL}
\end{align}
\end{subequations}

\noindent where $P_{\rm NL}$ is the non-linear power spectrum. Hence to generate these randoms walks, we need an estimate of both $P_{\rm lin}$ and $P_{\rm NL}$, which we obtain using CAMB~\citep{CAMB} with the fiducial cosmology of the DEUS N-body simulations ($\Lambda$CDM). 

\medskip
\noindent To compare the non-linear growth rate obtained from the MCRW with the one obtained from N-body simulations, we proceed as follows: we start by generating, at $a=1$, $100 000$ log-normal random walks,  that have ``physical" properties: for the overdense regions we require that $\Delta_{LN}>\Delta$ for $\Delta_{LN}>0$ and for the underdense regions if $\Delta_{LN}<0$ then $\Delta<0$. To obtain the profiles at higher redshift, we do not recompute all the walks at different redshifts, but we keep the values of all the linear trajectories at $a=1$,  $\Delta^{i}(a=1)$, where $i$, is the label of one selected random walk. We can therefore compute directly $\Delta^{i}(a)=\Delta^{i}(a=1) D^{+}(a)/D^{+}(a=1)$ (where $D^+(a)$ is the linear growth factor at $a$) and hence $\Delta_{\rm LN}^{i} (a)$ using Eq.~\ref{mapping}. From these profiles we compute the growth rate parameter $f$ using Eq.~\ref{fdef} and bin up the $f$ values according to $\Delta$, as we did for the simulation. Fig.~\ref{Fig2} shows the comparison of $f(\Delta)$ between the model and the simulation. Remarkably, even if the MCRW density profiles are not required to match the ones measured in the N-body simulation, the evolution of the non-linear growth rates as a function of the local density matches well between the model and simulations. The good agreement between our prediction with the N-body simulation  measurement suggests that it is possible to extract cosmological information from these non-linear regions.

\medskip
\noindent This is a key result that shows how the non-linear growth rate can be described by its local density. One can again draw an analogy with the island universe picture, where each region has its own growth rate depending on the mean density of the island. However when the size of the island is small, the coupling between small and large modes becomes complex. Hence the log-normal Monte Carlo Random Walks offer an alternative to model the environmental growth rate to extract cosmological information from these non-linear regions. Alternative method such as \cite{Bernardeau2015, Uhlemann2016, Codis2016} may also be useful to help improving the accuracy for the model prediction. 

\medskip 
\noindent To summarise, we have proposed an alternative approach to extract cosmological information from the non-linear regime. Instead of modelling ``out" the non-linear evolution of the growth rate down to a certain scale in the two point correlation function or power spectrum, aiming to recover the linear growth rate, we generalise $f$ in terms of local densities. This allow us to map the entire spectrum of the growth rate to its underlying cosmology. We have also shown as a proof of concept that the log-normal Monte Carlo Random Walk approach \cite{IA2016} describes the function of $f(\Delta)$ reasonably well. This in principle will allow us to extract cosmological information from measurement of $f(\Delta)$. 

\medskip
\noindent Futhermore, because our approach goes beyond Gaussian statistics (conventional RSD analysis use two-point statistics), we may expect to recover more information. We expect our approach to be particularly useful for testing theories of gravity which predict non-standard environmental dependence for structure growth. For example, in the $F(R)$ model, due to the chameleon screening mechanism, the strength of gravity differs in different local density \cite{Khoury2004, Carroll2005}. This may alter structure formation in a environmental dependent manner, which may be better captured by measuring $f({\Delta})$. Finally, the fact that the growth rate is lower/higher in voids/clusters than its linear version indicates that one need to employ non-linear modelling in these low/high density regions (e.g.~\citep{Cai16,IA2016,HamausSDSS,AchitouvPRD17,HamausSDSS2,Nadathur2017}) in order to have unbiased results. 

\medskip
\noindent The next question to ask is how to implement our method when analyzing galaxy surveys. The key is to be able to measure $f(\Delta)$ from data. We outline three possible approaches to do this. First, one can use galaxy coordinates to measure the galaxy number density $\Delta_g$ at different epochs. With a galaxy bias model, e.g. linear galaxy biasing, we can infer for $\Delta$  and use Eq.~\ref{fdef} to compute the growth rates. Second, with a combination of a galaxy redshift survey with a lensing survey, one can use the redshift survey data to define patches of over/under dense regions, and use the lensing survey to measure their $\Delta$'s in different tomographic bins. The recent work of \citep{Friedrich2017, Gruen17} have demonstrated the feasibility of this approach. Third, with the same survey set up as the second method, on top of measuring the densities with lensing for the patches defined in the galaxy field, one can perform $linear$ RSD analysis for the same patches in the redshift survey data. Indeed, keeping the island Universe analogy, the value of the growth rate derived using a simple multipole decomposition of the RSD should give us an estimate of the non-linear growth rate within those patches. We will investigate in more detail the implementation of our method in observations in future work. 

\medskip
\begin{acknowledgments}
\noindent \textbf{Acknowledgments}

\noindent We thank John Peacock, Masahiro Takada, Xin Wang  and Pengjie Zhang for useful discussions. The research leading to these results has received funding from the European Research Council under the European Community Seventh Framework Programme (FP7/2007-2013 Grant Agreement no.  279954) RC-StG \textit{EDECS}. YC was supported by supported by the European Research Council under grant numbers 670193.
\end{acknowledgments}
\bibliography{Achitouv-Cai}

\end{document}